\documentclass[aps,prl,showpacs,nobibnotes,twocolumn,
nofootinbib,nobalancelastpage,superscriptaddress]{revtex4}
\usepackage{graphicx}
\usepackage{amsmath}
\usepackage{epstopdf}
\usepackage{latexsym}
\usepackage{color}
\usepackage{dcolumn}% Align table columns on decimal point
\usepackage{bm}% bold math
\input{epsf}
\newcommand{\be}{\begin{equation}}
\newcommand{\ee}{\end{equation}}
\newcommand{\ba}{\begin{eqnarray}}
\newcommand{\ea}{\end{eqnarray}}
\newcommand{\bi}{\begin{itemize}}
\newcommand{\ei}{\end{itemize}}

\newcommand{\bfi}{\begin{figure}
\epsfxsize=9cm
\epsffile}
\newcommand{\efi}{\end{figure}}

\begin{document}
\title{Weighing the spatial and temporal fluctuations of the dark universe} 
\author{Pengjie Zhang}
\affiliation{Shanghai Astronomical Observatory, Chinese Academy of
  Science, 80 Nandan Road, Shanghai, China, 200030}
\affiliation{Joint Institute for Galaxies and Cosmology (JOINGC) of
SHAO and USTC, Shanghai, China}
\author{Rachel Bean}
\affiliation{Department of Astronomy, Space Sciences Building, Cornell University, Ithaca, NY 14853}
\author{Michele Liguori}
\affiliation{Department of Applied Mathematics and Theoretical
  Physics, Centre for Mathematical Sciences, University of Cambridge,
  Cambridge, CB3 0WA, United Kingdom}  
\author{Scott Dodelson}
\affiliation{Center for Particle Astrophysics,
Fermi National Accelerator Laboratory, Batavia, IL~~60510-0500}
\affiliation{Department of Astronomy \& Astrophysics, The University
  of Chicago, Chicago, IL~~60637-1433}
\email{pjzhang@shao.ac.cn,rbean@astro.cornell.edu,M.liguori@damtp.cam.ac.uk,dodelson@fnal.gov}     

\begin{abstract}
A generic prediction of the standard cosmology, based on general
relativity (GR), dark matter  and the cosmological constant (and more
generally, smooth dark energy),  is that, the two gravitational
potentials describing the spatial and temporal 
scalar perturbations of  the universe are equivalent. Modifications in GR or
dark energy clustering in general violate this relation. Thus this ratio serves
as a smoking  gun of the dark universe. We propose a method to
extract this ratio at various cosmological scales and redshifts from a set of
measurements, in a model independent way. The ratio measured by future surveys
has  strong discriminating power for a variety of dark universe scenarios. 
\end{abstract}
\pacs{98.80.-k;95.36.+x;04.50.Kd}
\maketitle
%%%%%%%%%%%%%%%%%%%%%%%%%%%%%%%%%%%%%%%%%%%

{ \bf Introduction}.---
Predictions based on general relativity (GR) plus the Standard Model of
particle physics  
are at odds with a variety of independent astronomical observations on
galactic and cosmological scales, implying failures in particle physics or
GR. There are various astrophysical tools to probe this dark side of
the physical universe (e.g. \cite{DETF,Jain07}). Combining them allows us
to break parameter degenaricies, reduce statistical errors and 
diagnose possible systematics. 

These multiple probes are also crucial to detect smoking guns of new
physics. For example, combining probes of the expansion history of the
universe and probes of the large scale structure,   the
relation between the expansion rate and structure growth rate can be checked
for signs of deviation from GR \cite{CR}.  Indeed, one of the key questions in
physics today is whether new particles/fields, such as dark matter and dark
energy, or modifications to GR are needed to explain the observations.

On large scales, two features of gravity can distinguish between a dark sector
and modified 
gravity~\cite{Zhang07,Amendola07,Caldwell07,Hu07,Bertschinger08}. One is the
effective 
Newton's constant $G_{\rm eff}$, which specifies the coupling between gravity and
matter. In GR, $G_{\rm eff}$ is equal to Newton's constant, but modified
gravity models often predict deviations. The other is the relation between the
two 
gravitational potentials $\phi$ and $\psi$. Here, the two potentials
are defined in  the Newtonian gauge through $\
ds^2=(1+2\psi)dt^2-a^2(1+2\phi)d{\bf x}^2$ where $a(t)$ is the scale
factor. The ratio $\eta\equiv -\phi/\psi$ 
weighs the relative ability of perturbations in matter-energy  to distort the
space-time.\footnote{Refer to other equivalent notations in
  \cite{Caldwell07,Amendola07,Bertschinger08}. An analogy of $\eta$ is the PPN
  parameter $\gamma$ (by forcing $\psi=-GM/r$ for point source). Solar system
  tests have revealed $\gamma=1\pm  O(10^{-5})$ \cite{SST} and provided strong
  support of GR. Constraints at galactic size and sub-cluster scales are
  consistent  with GR too \cite{gammaGC,Caldwell07}.} The standard cosmology,  based on GR, dark matter and the cosmological
constant (and more generally, smooth dark energy), predicts
$\eta=1$. Modifications from GR or emergence of intrinsic viscosity in  dark
energy fluid generally lead to $\eta$ deviating from unity.
Therefore, identifying observations, or
sets of observations, that will measure $G_{\rm eff}$ and $\eta$ is of
paramount
importance~\cite{Jain07,Zhang07,Caldwell07,Amendola07,Bertschinger08,Stabenau:2006td}.

In  \cite{Zhang07},  we showed how to isolate the first key feature,
feasibly testing 
the Poisson equation at $\sim 1\%$ accuracy level by combining weak lensing
with galaxy redshift distortion. 
In this paper, we will show that the same surveys allow us to directly measure
$\eta$, the second key feature,  at cosmological scales. This can be done in a
rather model independent manner. 

{\noindent\bf Models with $\eta\ne -1$}.---
Here we consider three models which produce deviations from the standard
prediction $\eta=1$ ($\phi+\psi=0$). 

Perturbations in the Dvali-Gabadadze-Porrati (DGP) model
\cite{DGP} have been carefully studied~\cite{DGPgamma,Koyama06}. For a flat DGP model, 
$\eta=[1-1/3\beta_{\rm  DGP}]/[1+1/3\beta_{\rm DGP}]$, where $\beta_{\rm
    DGP}=1-2r_cH(1+\dot{H}/3H^2)<0$ and $H$ is the Hubble expansion rate. Here
  $r_c$ is the cross-over scale beyond which higher dimensional effects become
  important. In a flat model with matter density $\Omega_m$,
  $r_c=1/H_0(1-\Omega_m)$. Since
$\beta_{\rm DGP}<0$, $\eta>1$ in this model and the deviation from unity can
be significant (Fig. \ref{fig:eta}). 
  
Another modified gravity model (which aims to eliminate dark matter, not dark
energy) is TeVeS \cite{TeVeS}, a relativistic version of
MOND\cite{MOND}. 
Besides the gravitational metric, TeVeS contains a scalar $\phi_S$ and a
vector field. 
It has been shown \cite{Skordis06,Dodelson06} that the TeVeS vector field 
can source the evolution of cosmological
perturbations and compensate for the lack of dark matter in the model. To fit
observations, the TeVeS parameter $K_B$ should be small, in which case the
vector perturbations  $\alpha$ and $E$ become
large. These vector perturbations then drive  $\eta$ to deviate from unity
\cite{Skordis06,Dodelson06,Schmidt07}, 
\be
\phi+\psi=e^{4\bar{\phi}_S} \left[
\dot{\zeta}+2\left(\frac{\dot{a}}{a}+2\dot{\bar{\phi}}_S\right)\zeta\right]
\ \ ,
\ee
 Here $\zeta\equiv (e^{-4\bar{\phi}_S}-1)\alpha$. Since the background value
 $\bar{\phi}_S\ll 1$ as 
 imposed by nucleosynthesis bounds, the deviation of $\eta$ from unity is
 mainly driven by the vector perturbation ($\phi+\psi\propto
 \bar{\phi}_S\alpha$).  A numerical evaluation of $\eta$ 
is shown in Fig. \ref{fig:eta}. For this figure we adopted a model
with $\Omega_b = 0.05$, $\Omega_\nu = 0.17$, $\Omega_{\Lambda}=0.78$ and no
dark matter.

A final possibility is that gravity is still GR, but dark energy
has non-negligible anisotropic stress $\sigma$ and causes inequality in
two potentials through \cite{Ma95} 
\be
\label{eqn:DE4}
\phi+\psi=-12\pi Ga^2(1+w)\bar{\rho}_{\rm DE}\ k^{-2}\sigma \ .
\ee
Although quintessence models predict $\sigma=0$, there are some dark energy
models that predict $\sigma\neq 0$ and $\eta\neq 1$ \cite{Hu99,Koivisto06}.  
%In
%general, intrinsic viscous forces could allow shear stresses to arise such
%that $\sigma\neq 0 $\eta\neq 1$ \cite{Hu99,Koivisto06}.
As a specific example, we consider an
extrinsic shear stress of the  form $\eta = 1/(1+\omega)$, with $\omega=
\omega_0a^3(1-\Omega_m)/\Omega_m$ with  
  $\omega_0$ constant, following  \cite{Caldwell07}.  In general,
  $\eta$ varies not only with time, but also with scale.  Richer physics
  encoded in the scale dependence of $\eta$ would allow  better
  discrimination between such dark energy model from other scenarios.

%%%%%%%%%%%%%%%%%%%%%%%%%%%%%%%%%
\bfi{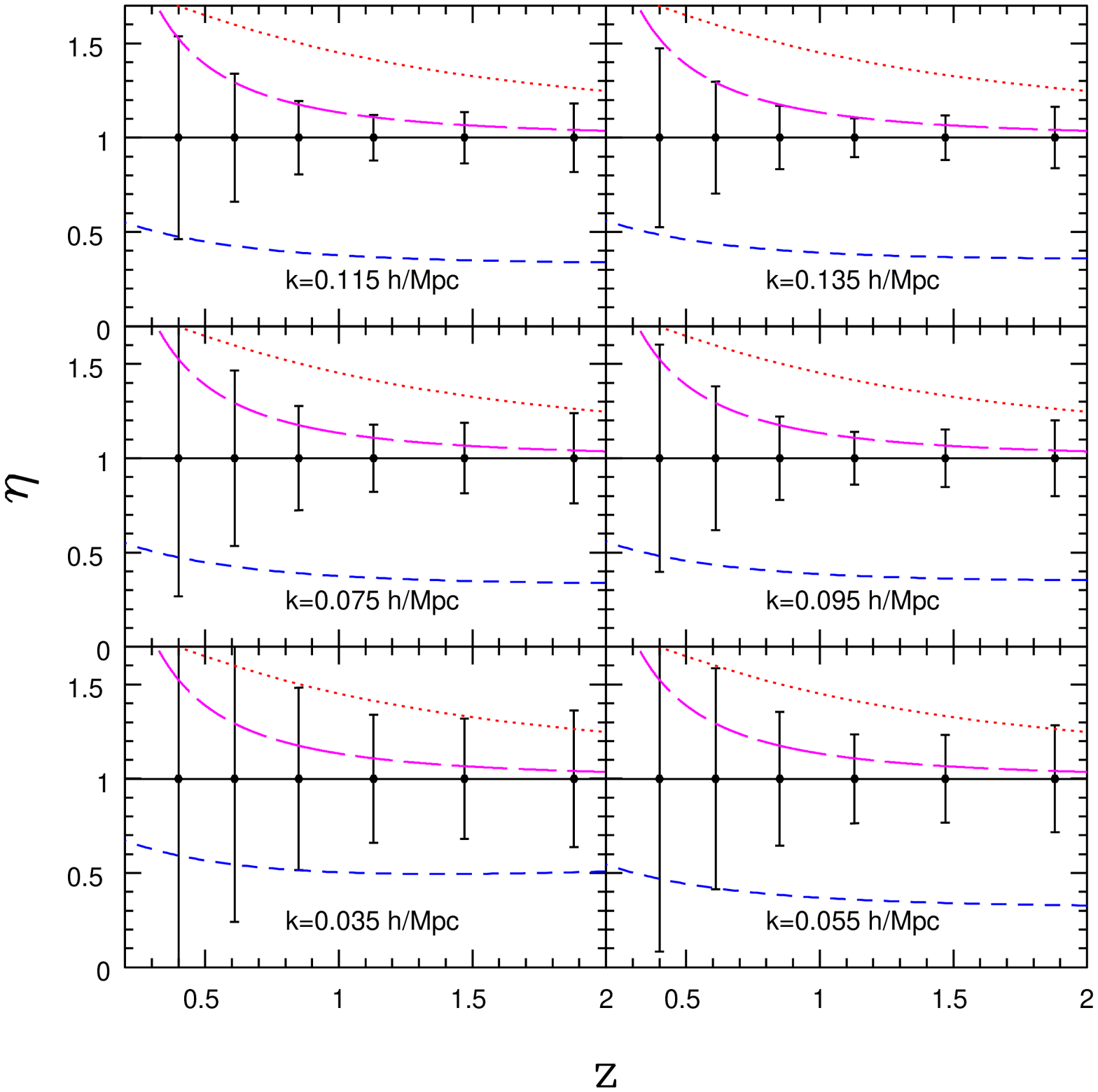}
\caption{Projected errors on $\eta$ from SKA (half sky coverage is
  adopted). Solid line is the
  prediction of the standard $\Lambda$CDM, on which the error forecast is
  based. We also show the predicted $\eta$ for a flat DGP cosmology
  with $\Omega_m=0.2$ (red, dotted line),  the anisotropic shear stress model
  with $\omega_0=-0.3$ (magenta, long dashed line) and TeVeS with 
  $K_B=0.08$  (blue, short dashed line). \label{fig:eta}}  
\efi
%%%%%%%%%%%%%%%%%%%%%%%%%%%%%%%
%%%%%%%%%%%%%%%%%%%%%%%%%%%%%%%%%%%%%%%%

{\noindent\bf The $\eta$ estimator}.--- To measure $\eta$, two independent
measures of 
gravitational potentials are required. Both $\nabla \psi$ and $\nabla
\phi$ source the particle acceleration. However, the contribution
from   $\phi$ is suppressed by a factor $v^2/c^2$, where $v$ is the particle
velocity. For this reason, non-relativistic particles such as galaxies only
respond to $\psi$.  For the same reason, photons respond equally to both the
potentials. Thus gravitational lensing measures the projected
$\nabla^2(\phi-\psi)$   along  the line of sight. 
We propose an estimator consisting of the cross-correlation of each (the lensing
field and the velocity field) with the galaxy distribution. 

The first cross-correlation is the lensing measurement with galaxy
over-density in a 
narrow redshift bin\cite{Zhang07}. We can then obtain the cross-power
spectrum  $P_{\nabla^2(\phi-\psi)g}(k,z)$ between  $\nabla^2(\phi-\psi)$ and the
galaxy number overdensity in the redshift bin associated with the galaxies. 

The second cross-correlation power spectrum $P_{\theta g}$ can be obtained
from the redshift distortions of the 
galaxy distribution in a spectroscopic
survey~\cite{Tegmark02,Zhang07,Jain07,Zhang08}. Here, 
$\theta_g\equiv -\nabla\cdot 
  \vec{v}_g/H$ and $ \vec{v}_g$ is the comoving peculiar velocity. 
We show below that this cross-spectrum is directly related to $P_{\nabla^2\psi g}$,
but first let us assume that this is so, that $P_{\nabla^2\psi g}$ can be extracted
from the $\theta$-$g$ cross-correlation. In that case, 
the ratio of these two cross-spectra leads to an
estimator for $\eta$: 
\be
\label{eqn:eta}
\hat\eta=\frac{P_{\nabla^2(\psi-\phi) g}}{P_{\nabla^2\psi g}}-1\ .
\ee

To see that $\theta$ is related to $\psi$, recall that on large scales
gravity is the only force 
accelerating galaxies, so  $d(a\vec{v}^p_g)/dt=-\nabla \psi$, where
$\vec{v}^p_g=a\vec{v}_g$ is the proper motion. Taking the divergence of this
leads to 
\be
\label{eqn:psi_theta}
\nabla^2\psi=-\frac{d(a^2\nabla\cdot
  \vec{v}_g)}{dt}=-\left(\ln[a^2HD_{\theta}]\right)^{'}a^3H\nabla\cdot \vec{v}_g (a)\ .
\ee
Here, $^{'}\equiv d/da$ and $D_{\theta}$ is the growth factor of
$\theta_g$. The last relation holds in the linear regime where different modes decouple. We then have 
\be 
\label{eqn:pst}
P_{\nabla^2\psi g}=-\left(\ln[ 
a^2HD_{\theta}]\right)^{'} a^3H^2 P_{\theta g},
\ee 
the desired relation.  
 
The proportionality factor relating the two cross-spectra in
Eq.~(\ref{eqn:pst}) requires knowledge of the expansion rate $H(z)$ and the
growth factor $D_\theta(z)$. We assume that the former can be measured by
other means; indeed our goal is to distinguish dark sector models which
produce identical expansion histories. No such assumption is needed for the
growth factor, because the same survey that measures $P_{\theta g}$ will also
measure $P_{\theta\theta}$,
which is proportional to $D_{\theta}^2$ and thus measurement of $P_{\theta \theta}$ in multiple
redshift bins can be used to recover $(\ln 
a^2HD_{\theta})^{'}$  (see the appendix for details). We adopt the
minimum variance estimator to 
estimate errors in the reconstruction of $P_{\theta g}$ and $P_{\theta
  \theta}$ \cite{Zhang07,Zhang08}.  This reconstruction adopts no assumption
on galaxy bias, so it is less affected by possible stochasticity or scale
dependence in galaxy bias.

Application of the $\eta$ estimator in Eq.~(\ref{eqn:eta}) relies on the
condition of  linear evolution such that  
Eq.~(\ref{eqn:psi_theta}) and therefore (\ref{eqn:pst}) hold. For this reason,
we restrict our discussion to the linear regime. This approach
is robust against several  
uncertainties: (1) It does not suffer uncertainty induced by the
galaxy bias, whose effect cancels when taking the ratio in Eq.~(\ref{eqn:eta}).  (2) It is not 
susceptible to possible galaxy velocity bias, defined with respect to
peculiar velocity  of dark  matter or dark energy, since we directly
measure $(\ln a^2HD_{\theta})^{'}$, instead of relying on a theory to
calculate it. (3) It is applicable to general dark energy models and
modified gravity models. It does not require dark energy to be smooth,
nor gravity to be minimally coupled, nor scale-independent
$D_{\theta}$.

{\noindent\bf Forecast}.---In order to measure $\eta$ in this way, the lensing
and 
redshift surveys must be sufficiently deep and wide. The proposed
spectroscopic galaxy 
survey ADEPT or 21cm survey HSHS
\cite{Peterson06}, combined with a lensing survey such as LSST, would
be sufficient. Alternatively, SKA
%\footnote{http://www.skatelescope.org} 
alone would be 
able to provide both suitable lensing, through cosmic shear \cite{Blake04} and
cosmic magnification \cite{Zhang05}, and  galaxy redshift measurements, as
potentially would  the
Euclid\footnote{http://sci.esa.int/science-e/www/area/index.cfm?fareaid=102}
mission. So  we focus on SKA projections.  $D_{\theta}$ can be
measured by SKA at multiple bins of redshift and scale to impressive accuracy
(Fig. \ref{fig:Dtheta}). We then infer $(\ln[a^2HD_{\theta}])^{'}$ from the
above measurements.

Projections for the errors on $\eta$ from SKA in a variety of $(k,z)$ bins are
shown in Fig.~\ref{fig:eta}. One example of the power of this measurement is
in constraining the DGP model. The $E_G$ measurement proposed
in~\cite{Zhang07} can only marginally distinguish the $\Omega_m=0.2$ flat DGP
model from  $\Lambda$CDM. Fig.~\ref{fig:eta} shows, though, that
these models have 
significantly different predictions for $\eta$; The TeVeS model
adopted has been shown to produce a good fit of CMB and LSS data 
\cite{Skordis06}. However, with large
deviation from $\eta=1$, this model can be unambiguously distinguished from
$\Lambda$CDM.  Thus $\eta$ and $E_G$ are
highly complementary to probe the dark universe\footnote{Errors in
  $\eta$ and in $E_G$ are partly  
correlated. Future work should take this into account by fitting $\eta$ and
$E_G$ simultaneously, while marginalizing all other parameters. };
Modifications in gravity or dark energy viscosity often lead to stronger
scale dependence in $\eta$ than what is shown in  Fig.~\ref{fig:eta}. Our $\eta$
estimator could have stronger 
discriminating  power for these models.

%%%%%%%%%%%%%%%%%%%%%%%%%%%%%%%%%%%
\bfi{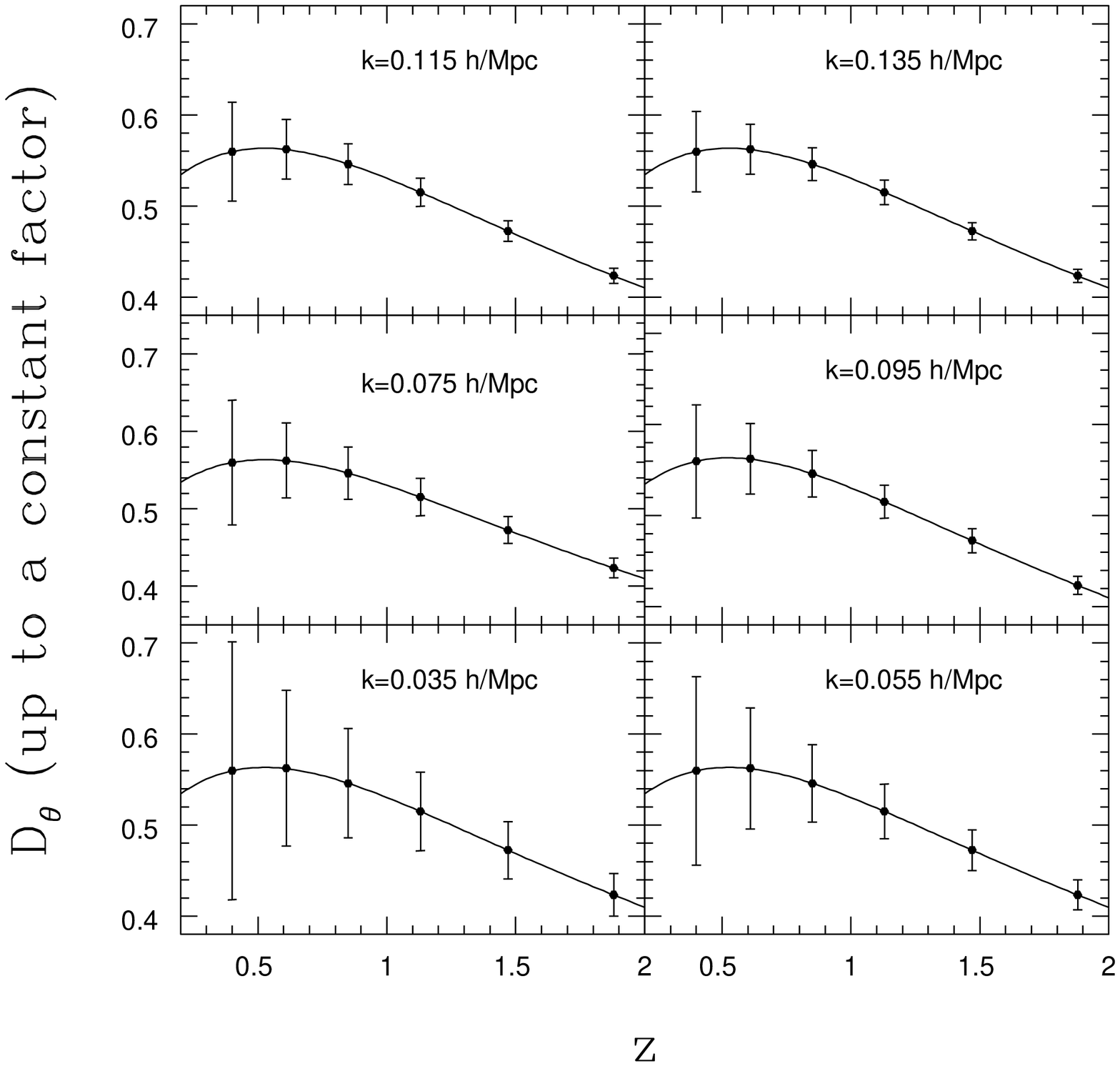}
\caption{Projected errors on $D_{\theta}$ from SKA. Since SKA measures $P_{\theta \theta}$, which is proportional to
  $D_{\theta}^2$, this tells us directly on $D_{\theta}$, up  to a
  normalization, which is of no physical 
  importance in $\eta$ measurement. The fiducial cosmology is the standard
  $\Lambda$CDM cosmology (solid lines). The bin size is $\Delta k=0.02
  h/$Mpc. For any realistic surveys, we  have only limited ${\bf k}$ sampling.
At radial direction, the  available modes are
$k_z=2\pi H(z) i/\Delta z$, where $i=1,2,\cdots$. This is the ultimate
  limiting factor of ${\bf k}$ sampling in wide  field surveys such as SKA. 
Since we need to compare different redshifts to obtain
$\ln(a^2D_{\theta})^{'}$, we require the ${\bf k}$ sampling at relevant
redshift bins to be roughly identical.  This requires $\Delta 
z_{i+1}/H(z_{i+1})=\Delta z_i/H(z_i)$. We choose $(z_1,\Delta
z_1)=(0.4,0.2)$. Thus $(z_2,\Delta z_2)=(0.61,0.22)$, $(z_3,\Delta
z_3)=(0.85,0.26)$,$[z_4,\Delta z_4]=[1.13,0.31]$, $ [z_5,\Delta
  z_5]=[1.47,0.37]$, $[z_6,\Delta z_6]=[1.88,0.45]$. 
The maximum  matter density power spectrum
variance is max[$\Delta^2(k,z)]<0.3$ for the $k$ and $z$ bins shown, allowing us to  neglect the effect of
non-linearity for the moment.  \label{fig:Dtheta}} 
\efi
%%%%%%%%%%%%%%%%%%%%%%%%%%%%%%

There is room to  improve the $\eta$ measurement. (1) The fractional  
error in $P_{\theta\theta}$ reconstructed from the redshift distortion is
about 15 times larger than the cosmic variance limit \cite{Zhang08}. Adopting
the  approximation of deterministic galaxy bias, the associated error will
decrease 
by a factor of 3. However, to reach the cosmic variance limit, other velocity
measurement techniques should be explored (e.g. \cite{Zhang08b}). (2) The forecast outlined above only uses $D_{\theta}$ measurements from
$z<2$  galaxies and thus  limited the accuracy of  $(\ln
a^2HD_{\theta})^{'}$ measurement. Depending on the design and on the nature of
21cm emitting galaxies,  SKA 
may allow measurements of $D_{\theta}$ at higher redshifts. Furthermore,
$D_{\theta}$ at even higher redshifts ($z\sim 10$) can be measured from
redshift distortions of diffuse 21cm background. Improvement in
the measurements of $(\ln a^2HD_{\theta})^{'}$ and $\eta$ at $z<2$ would
result  from  the inclusion of such observations. (3) Furthermore, measurements
of  $\eta$ at   $z>2$ can be made feasible by the inclusion of CMB lensing and 21cm
background lensing. 

%%%%%%%%%%%%%%%%%%%%%%%%%%%%%%%%%%%%%%%%%%%
\medskip 
We have shown that future precision imaging surveys of weak gravitational
lensing and spectroscopic 
surveys of galaxy redshift distortions provide highly 
complementary methods to probe the dark universe. In combination they allow us to
isolate two key features of the dark universe, the effective Newton's constant
$G_{\rm eff}$ and $\eta\equiv -\phi/\psi$, from many astrophysical
complexities, and distinguish competing scenarios of the dark universe
robustly.

%%%%%%%%%%%%%%%%%%%%%%%%%%%%%%%%%%%%%%%%%%%
\medskip 
{\bf Acknowledgments.}--- PJZ is supported by  the National Science
Foundation of China  grant 10533030, 10673022, CAS grant KJCX3-SYW-N2 and the
973 program grant No. 
2007CB815401. RB's work is
supported by NASA ATP grant NNX08AH27G, NSF grants AST-0607018 and PHY-0555216
and Research Corporation. SD is supported by the US  Department of Energy.

%%%%%%%%%%%%%%%%%%%%%%%%%%%%%%%%%%%%%%%%%%%
\medskip 
{\bf Appendix.}---To infer $(\ln a^2HD_{\theta})^{'}$ from $D_{\theta}$
measured in limited redshift bins, a parametrization of
$D_{\theta}$ is required.  Since $D_{\theta}$  evolve smoothly, 
 $(\ln a^2HD_{\theta})^{'}$ should not be strongly dependent on the precise
 form of the 
parametrization. In this paper, we extend a widely used parameterization for
$D_{\theta}$ in standard gravity. For gravity models minimally coupled to
matter, 
$D_{\theta}=D^{'}a=fD$, where $f\equiv d\ln D/d\ln a$ and $D$ is the
linear density growth factor.  One approximation adopted in the literature is
$f\simeq (\Omega_ma^{-3}/E^2)^{\gamma}$ (e.g. \cite{MMG}). Here,  $E\equiv
H/H_0$ is the normalized Hubble parameter. This approximation works well
not only for  CDM ($\gamma=5/9$ for $\Omega_{\Lambda}+\Omega_m=1$ and
$\gamma=0.6$ for $\Omega_{\Lambda}=0$), but also for some  modified gravity
models 
such as DGP ($\gamma=2/3$, \cite{DGPgamma}).    We thus propose to fit a
parameterization 
\be
f_*\equiv \left(\frac{\Omega_*a^{-3}}{E^2}\right)^{\gamma_*}\ .
\ee
Here, both $\Omega_*$ and $\gamma_*$ are parameters to be fitted for each $k$
bin. $D$ and $D_{\theta}$ are then obtained by the relation $f\equiv d\ln D/d\ln
a$ and $D_{\theta}=fD$.

%%%%%%%%%%%%%%%%%%%%%%%%%%%%%%%%%%%%%%%%%%%

\end{document}